\documentclass[amsmath,amssymb,notitlepage, twocolumn,aps,pra,floatfix]{revtex4-2}
\usepackage{xcolor}
\usepackage[colorlinks=true, allcolors = black]{hyperref}
\usepackage{url}
\usepackage{qcircuit}
\usepackage{subfigure}
\usepackage{graphicx}
\usepackage{dcolumn}
\usepackage{bm}

\usepackage{multirow}
\usepackage{braket}
\usepackage{amsmath}
\usepackage{cleveref}
\usepackage{mathtools}
\DeclarePairedDelimiter\ceil{\lceil}{\rceil}

\newcommand{\abs}[1]{\lvert#1\rvert}

\begin{document}
\preprint{APS/123-QED}

\title{Preparing Dicke states in a spin ensemble using phase estimation\\}

\author{Yang Wang}
\affiliation{QuTech, Delft University of Technology, P.O. Box 5046, 2600 GA Delft, The Netherlands}
\author{Barbara M. Terhal}
 \affiliation{QuTech, Delft University of Technology, P.O. Box 5046, 2600 GA Delft, The Netherlands and JARA Institute for Quantum Information, Forschungszentrum Juelich, D-52425 Juelich, Germany}
\date{\today}

\begin{abstract}
We present a Dicke state preparation scheme which uses global control of $N$ spin qubits: our scheme is based on the standard phase estimation algorithm, which estimates the eigenvalue of a unitary operator. The scheme prepares a Dicke state non-deterministically by collectively coupling the spins to an ancilla qubit via a $ZZ$-interaction, using $\ceil*{\log_2 N} + 1$ ancilla qubit measurements.  The preparation of such Dicke states can be useful if the spins in the ensemble are used for magnetic sensing: we discuss a possible realization using an ensemble of electronic spins located at diamond nitrogen-vacancy centers coupled to a single superconducting flux qubit. We also analyze the effect of noise and limitations in our scheme.
\end{abstract}
\maketitle
\section{Introduction}
A promising application of the emerging quantum technology is quantum-enhanced sensing, sometimes referred to as quantum metrology \cite{degen2017quantum, toth2014quantum}.
Using entangled states, one can, in principle, improve the measurement sensitivity from the standard quantum limit ($1/\sqrt{N}$) to the Heisenberg limit ($1/N$) \cite{giovannetti2004quantum,giovannetti2011advances,degen2017quantum}, where $N$ is the number of probes or repetitions. However, preserving this quantum advantage is difficult in the presence of decoherence \cite{demkowicz2012elusive}. For instance, a single-qubit Pauli $Z$-error can totally dephase a $N$-qubit GHZ-state, which would obtain Heisenberg-limited sensitivity in the noiseless case \cite{toth2012multipartite}.

$N$-qubit Dicke states form a class of entangled states which are interesting for metrology \cite{hakoshima2020efficient, pezze2018quantum, zhang2014DickeSqueezed, apellaniz2015DcikeUsefulDetecting, toth2012multipartite}. Compared to other states used in quantum sensing, Dicke states have been argued to be more robust to various noise sources such as spin dephasing, spin damping, and spin number fluctuations \cite{zhang2014DickeSqueezed}. Recent work has demonstrated a scheme to use Dicke states for detecting the magnetic field induced by a single spin \cite{hakoshima2020efficient}. Another distinctive feature of the use of Dicke states is that  the optimal sensitivity can be obtained through only global control on the set of spins \cite{apellaniz2015DcikeUsefulDetecting, zhang2014DickeSqueezed}. This is relevant for realizing practical quantum sensing using entangled states, as precise individual spin qubit control can be difficult. Furthermore, superpositions of Dicke states can be used for quantum error correction \cite{ouyang2014permutation, ouyang2019robust}.

Dicke state preparation has been experimentally realized using photons \cite{DickeFourPhoton,DickeSixPhoton} and trapped-ion qubits \cite{DickeAdiabaticPassage, DickeIonChain}, and there also exist many theoretical preparation proposals suitable for a few qubits, see Ref.~\cite{DickUltrastrong, DickeSinglePhotonDetection, DickeSelectiveInteraction,GeneralizedParity} for example. However, it remains a challenge for large spin ensembles like $N> O(100)$ diamond Nitrogen-Vacancy centers (negatively-charged NV \cite{mohamed2021NV}), each hosting an electronic spin $S=1$. Since these NV center spins are rather isolated from each other, it is costly to perform entangling gates between the electronic spins \cite{dolde2014high,dolde2013room,neumann2010quantum}. 
This limitation excludes quantum algorithms for preparing Dicke states which are based on the full addressability of the qubits \cite{bartschi2019deterministic,mukherjee2020preparing}. 
To address this issue, some work has been dedicated to schemes which require only a global control of the spin ensemble, such as using steady state evolution \cite{masson2019extreme}, repeated energy transfer \cite{hakoshima2020efficient}, continuous weak measurements \cite{DickeContinuousMeasure}, and the use of geometric phase gates \cite{johnsson2019geometric}. Unfortunately, these methods are still demanding currently when $N$ is large, as they often need complicated measurement-based feedback, high fidelity control and long preparation times. For example, the optimized scheme in Ref.~\cite{hakoshima2020efficient} uses $O(N)$ rounds of initialization and evolution of an ancilla qubit. 
Our goal is to improve the scaling with $N$ so that one could possibly handle a larger error rate on the ancilla qubit.

In this paper, we present a Dicke state preparation scheme that uses standard phase estimation \cite{nielsen2002quantum}, which prepares an eigenstate of a unitary operator by estimating its eigenvalue. This algorithm is based on executing projective measurements on the spin ensemble using an ancilla qubit, and it will prepare a random Dicke state. 
The scheme requires a $ZZ$-coupling between each spin in the ensemble to a single ancilla qubit. In Section \ref{sec:systemHamiltonian} we detail how this coupling could be realized between an ensemble of NV electronic spins and a superconducting flux qubit as ancilla. 

Our scheme is efficient with respect to the number of operations.
It uses only $\ceil*{\log_2 N} + 1$ rounds of phase estimation for preparing a random $N$-spin Dicke state. 
Each round of phase estimation measures a global operator of the spins, i.e., it applies an ancilla-qubit controlled global $J_z=\frac{1}{2}\sum_{i=1}^N Z_i$ rotation followed by ancilla qubit readout. The total time for performing the controlled rotations is upper-bounded by a constant and the preparation time thus scales as $O(\log_2 N)$. With a probability $\sim O(1/\sqrt{N})$, the prepared Dicke state would obtain Heisenberg-limited sensitivity using only global control.

Besides the efficiency, our scheme also has some  noise-resilience: phase estimation can be realized with integrated dynamical decoupling, which provides robustness to the dephasing of the ancilla qubit as well as the dephasing of the spins in the ensemble. Furthermore, by repeating the projective measurements and performing a simple majority vote, the effects of ancilla qubit decay and flipped measurements (due to ancilla qubit dephasing or imperfect measurement) can be mitigated.  

This paper is organized as follows. In Sec.~\ref{sec:DickeDefinition}, we briefly review Dicke states and Heisenberg-limited sensing. In Sec.~\ref{sec:IdealPreparation} we present the idea of using phase estimation to prepare Dicke states.  In Sec.~\ref{sec:systemHamiltonian} we discuss the Hamiltonian and a possible experimental setup with multiple NV centers coupled to a flux qubit. 
In Sec.~\ref{sec:noisyPreparation} we numerically consider the performance of the scheme given the dominant noise sources. Finally, we discuss the results in Sec.~\ref{sec:Discussion}.

\subsection{Dicke states}\label{sec:DickeDefinition}

For simplicity, we assume even spin number $N$ throughout this paper (odd spin-number $N$ can be treated similarly). The $N$-spin (or qubit) Dicke state $\ket{N, m_z}$ with $m_z \in \{-\frac{N}{2}, \ldots, \frac{N}{2}\}$ is a uniform, permutation-symmetric, superposition of $N$-bit strings $\ket{x}$ where all bit-strings have $N/2 + m_z$ spins in $\ket{0}$, i.e., their Hamming weight is $N/2-m_z$.
For example,  $\ket{N=4, m_z = 0} = \frac{1}{\sqrt{6}}(\ket{0011} + \ket{0101} + \ket{0110} + \ket{1001} + \ket{1010} + \ket{1100})$. A Dicke state $\ket{N,m_z}$ is an eigenstate of the collective spin operator 
\begin{equation}
 J_z = \frac{1}{2} \sum_{i=1}^N Z_i,   
 \label{eq:defJz}
\end{equation} with eigenvalue $m_z$. Here $Z_i$ is the Pauli $Z$ operator on the spin labeled $i$. In addition, we have $J_x=\frac{1}{2}\sum_{i=1}^N X_i $ and $J_y=\frac{1}{2}\sum_{i=1}^N Y_i$.

To use such states for metrology, one imagines that the prepared quantum state is transformed by $e^{-i\theta J_y}$ and the goal is to estimate the rotation angle $\theta$ which is assumed to be small. 
A standard metrological method (for NV centers, limited by $T_2$ and optical measurement accuracy) is Ramsey spectroscopy \cite{barry2020sensitivity} using a single qubit state repeatedly (or, equivalently, using a product state of multiple qubits). In this context, the Ramsey method corresponds to preparing a simple product state $e^{i \frac{\pi}{2} J_y}\ket{00\ldots 0}$ and letting it thus evolve to $e^{-i (\theta-\frac{\pi}{2}) J_y}\ket{00\ldots 0}=(\frac{1}{\sqrt{2}}(\ket{+}_Y+e^{i(\theta-\frac{\pi}{2})}\ket{-}_Y)^{\otimes N}$. The rotation angle $\theta$ can then be estimated by measuring each spin in $Z$-basis. The measurements give the expectation value $\langle J_z(\theta)\rangle=\frac{N}{2} \sin(\theta)$, which is most sensitive to small perturbations of $\theta$ around $\theta = 0$ \cite{degen2017quantum}. The sensitivity of a product state is limited by the standard quantum limit, i.e., the variance in $\theta$ scales as $(\Delta \theta)^2 \sim 1/N$. It has been argued that Dicke states for $m_z=O(1)$ can reach the Heisenberg-limited sensitivity (i.e., $(\Delta \theta)^2 \sim 1/N^2$), as follows.

In general, one will measure some operator $\mathcal{M}$ on the final state $\exp(-i\theta J_y) \ket{N,m_z}$ to estimate the value of $\theta$. The variance of $\theta$ can be calculated by the error propagation formula 
\begin{equation}
    (\Delta \theta)^2 = \frac{(\Delta \mathcal{M}(\theta))^2}{\abs{\partial_{\theta} \braket{\mathcal{M}(\theta) } }^2},
\end{equation}
where the expectation value is with respect to the initial state $\ket{N,m_z}$ and $\mathcal{M}(\theta)$ is the Heisenberg-evolved operator. If we were to measure $\mathcal{M}=\alpha J_x+\beta J_z$, then 
\begin{equation}
\begin{split}
    \braket{\mathcal{M}(\theta)} &= \braket{J_z}(\beta \cos(\theta)+\alpha \sin(\theta)) \\
    & \approx m_z (\beta+\theta \alpha),
\end{split}
\end{equation}
for small $\theta$ (note that $\bra{N,m_z} J_x \ket{N,m_z}=0$). We measure $J_x$ by choosing  $\beta=0, \alpha=1$, its expectation value $\braket{J_x(\theta)}$ has an optimal dependence on $\theta$ when $m_z$ is large. However, the variance $(\Delta J_x(\theta))^2$ will be large in a rotated Dicke state, precluding any Heisenberg gains. 

The proposal is instead to measure $\mathcal{M}=J_z^2$, so that the variance is given by (see details in Ref.~\cite{apellaniz2015DcikeUsefulDetecting}):
\begin{equation}
\begin{split}
    (\Delta \theta)^2 & = [(\Delta J_x^2)^2 f(\theta)+4\langle J_x^2\rangle-3\langle J_y^2\rangle  \\
& \!\! -2\langle J_z^2\rangle \times(1+\langle J_x^2\rangle)+6\langle J_z J_x^2 J_z\rangle ]\,/\, [4(\langle J_x^2\rangle-\langle J_z^2\rangle)^2]
\end{split}
\end{equation}
with $f(\theta)=\frac{(\Delta J_z^2)^2}{(\Delta J_x^2)^2\tan^2(\theta)}+\tan^2(\theta)$. The minimal variance is obtained when 
$\tan^2(\theta) = \sqrt{(\Delta J_z^2)^2/(\Delta J_x^2)^2}$. For Dicke state $\ket{N,m_z}$ the minimal variance (obtained at $\theta \approx 0$) is
\begin{align}
    (\Delta \theta_{\rm min})^2 = \frac{2m_z^2+2}{N^2+2N-12m_z^2} + \frac{64m_z^4 -16m_z^2}{(N^2+2N-12m_z^2)^2}.
\end{align}
Note that the sensitivity can surpass the standard quantum limit when $m_z \sim O(\sqrt{N})$ and is Heisenberg-limited when $m_z \sim O(1)$. In addition, when $m_z = 0$, $(\Delta \theta_{\rm min})^2 = \frac{2}{N(N+2)}$ saturates the quantum Cram\'er-Rao bound \cite{toth2012multipartite}.
The expectation value $\braket{J_z^2}$ can in principle be obtained by measuring $J_z$, squaring its outcome and gathering sufficient statistics by repeating the measurements.
We are thus especially interested in Dicke states close to $\ket{N,0}$, i.e., $\ket{N,m_z}$ with $m_z \sim O(1)$. Other than this motivation, we do not focus on aspects of using a (noisy) Dicke state for metrology in this paper.

\section{Phase estimation preparation for Dicke states}\label{sec:IdealPreparation}

In this section, we will show how to prepare a Dicke state using a phase estimation algorithm. 

Phase estimation of a unitary operator is the process of measuring its eigenvalue and simultaneously projecting the input state to the corresponding eigenstate. This idea has for example been proposed to prepare Gottesman-Kitaev-Preskill states in a bosonic system, realized by determining the eigenvalues of two unitary operators approximately \cite{duivenvoorden2017single, terhal2016encoding}. 

For preparing Dicke states, we will start from a product state, e.g. Eq. \eqref{eq:initial_distribution}, where all spins in the ensemble are in the same state. Such a product state is clearly already permutation-symmetric but not yet an eigenstate of $J_z$. Since the Dicke state $\ket{N, m_z}$ is the unique $N$-qubit permutation-symmetric eigenstate of the operator $J_z$ with eigenvalue $m_z$, we can then prepare a Dicke state via phase estimation. This is realized by measuring the eigenvalues of a unitary operator $U$ whose eigenvalues are in 1-1 correspondence to the eigenvalues $m_z$. Note that it is important to start the phase estimation scheme in the permutation-symmetric subspace, as $J_z$ has eigenstates outside of this permutation-symmetric subspace on which we do not want to project.

Since the eigenvalue $m_z \in [-N/2, N/2]$, the integer $m_z + 2^K$ with $K = \ceil*{\log_2 N}+1$ is positive.
To find the unitary operator for phase estimation, we write down the binary representation
\begin{align}\label{eq:binary_rep}
    m_z + 2^K = \sum_{l=1}^{K+1} b_l 2^{l-1}.
\end{align}
Note that the value of $m_z$ can be unambiguously determined using the first $K$ of $K+1$ bits (i.e. $b_l= 0,1$ with $l=1,2,\ldots K$).
Then the unitary operator for phase estimation is
\begin{equation}\label{eq:pe_unitary}
    U = e^{i2\pi(J_z+2^K)/2^K} = e^{i2\pi J_z/2^K}.
\end{equation}
This gives $U\ket{N, m_z} = e^{i\phi(m_z)}\ket{N,m_z}$, where $\phi(m_z)= \pi\sum_{l=1}^{K} b_l 2^{l-K}$ is indeed an 1-1 function of the first $K$ bits in Eq.~(\ref{eq:binary_rep}).  Therefore, the preparation of a Dicke state is transformed to the task of performing phase estimation for this unitary operator $U$.

\begin{figure}
	\[
	\Qcircuit @C=0.6em @R=0.6em{
    &&&&&&\lstick{\text{spins:}\, \ket{\psi_{j-1}}} & \qw {/} &\qw & \gate{U^{2^{K-j}}} & \qw
    & \qw \\
  &&&&&&\lstick{\text{ancilla qubit:}\, \ket{0} } & \gate{\mathrm{H}} & \qw & \ctrl{-1}  &\gate{R_z(\vartheta) }&\gate{\mathrm{H}} &\meter
	}
	\]
	\caption{ The $j$th round phase estimation for the unitary operator $U = e^{i2\pi J_z/2^K}$ in Eq.~(\ref{eq:pe_unitary}). This circuit projectively measures the eigenvalues of the unitary operator $U_j = e^{i\pi 2^{1-j} (J_z - A_{j-1})}$ on the input state $\ket{\psi_{j-1}}$ in Eq.~(\ref{eq:jthOutput}). Before the measurement, the ancilla qubit is rotated around the $Z$-axis by the angle $\vartheta = \pi A_{j-1}2^{1-j}$, with $A_{j} = \sum_{l=1}^{j} 2^{l-1} b_{l}$. Here $b_l = 0,1$ is the measurement outcome of the previous measurement of $U_l$. }
	\label{fig:pe_DickePreparation}
\end{figure}
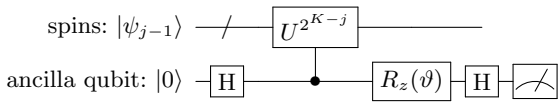

Using phase estimation, one cannot prepare a specific Dicke state $\ket{N,m_z}$ deterministically, as there is in general no easy operation that could transform $\ket{N,m_z}$ to $\ket{N,m_z'\neq m_z}$ \cite{kobayashi2014DickeTransform}. However, we can easily maximize the probability of obtaining a Dicke state whose sensitivity is Heisenberg-limited. This requires starting from the product state
\begin{align}\label{eq:initial_distribution}
    \ket{\psi_0} =  \left(\frac{\ket{0}+ \ket{1}}{\sqrt{2}}\right)^{\otimes N} = \sum_{m_z=-N/2}^{N/2} \sqrt{p(m_z)}\ket{N,m_z},
\end{align}
where $p(m_z)$ is a binomial distribution with average $\langle m_z \rangle=0$ and standard deviation $\sqrt{N}/2$ (i.e., $p(m_z) = \binom{N}{m_z+N/2}/2^N$).  This distribution reaches its maximum at $m_z =0$ and $p(m_z=0)\approx \sqrt{2/(\pi N)}$ (using Stirling's approximation). Dicke states $\ket{N,m_z}$ with $m_z \sim O(1)$ can thus be obtained with a probability $O(1/\sqrt{N})$. To prepare these states, one would thus need to repeat the preparation $O(\sqrt{N})$ times on average. 

Among many other variants \cite{terhal2016encoding,O_Brien_2019}, we choose standard or `textbook' phase estimation: standard phase estimation uses only $K$ measurements to determine the eigenvalue of $J_z$ by determining the first $K$ bits in Eq.~(\ref{eq:binary_rep}). Furthermore, these measurements can be executed in a sequential manner, where only one ancilla qubit is required. The ancilla qubit is used as the control to apply controlled-$U^{2^{K-j}}$ gates with $j = 1,2 \ldots, K$ starting at $j= 1$, for which $U^{2^{K-1}}=\exp(i \pi J_z)$.

The circuit of the $j$th round phase estimation
is shown in Fig.~\ref{fig:pe_DickePreparation}, where the ancilla qubit is measured in a basis determined by previous measurement outcomes $m_i =0, 1$ with $i=1,2,\ldots j-1$. Before readout, the ancilla qubit is rotated around the $Z$-axis by the angle $\vartheta = \pi A_{j-1}2^{1-j}$, where $A_{j} = \sum_{l=1}^{j} 2^{l-1} b_{l}$ (and $A_0 = 0$). The $j$th round phase estimation is described by the projector
\begin{equation}\label{eq:idealprojector}
\begin{split}
    \mathrm{P}(b_j) &= \frac{1 + (-1)^{b_j} U_j}{2},\\
    U_j &= e^{i\pi 2^{1-j} (J_z - A_{j-1})},
\end{split}
\end{equation}
We note that $\mathrm{P}(b_j=0)\mathrm{P}(b_j=1)=0$ as $U_j$ has eigenvalues $\pm 1$ on the space of states with given value for $A_{j-1}$. After $j$ rounds of phase estimation, the spins in the ensemble are projected into a superposition of Dicke states, i.e.,
\begin{equation}\label{eq:jthOutput}
\begin{split}
    \ket{\psi_{j}} &= \frac{1}{\sqrt{\mathcal{N}_j}} \mathrm{P}(b_j)\cdots \mathrm{P}(b_2)\mathrm{P}(b_1)\ket{\psi_0}\\
    &= \frac{1}{\sqrt{\mathcal{N}_j}}  \sum_{n \in \mathrm{Z}} \sqrt{ p(2^j n + A_j)  }\ket{N,2^j n + A_j},
\end{split}
\end{equation}
where $\mathcal{N}_j$ is the normalization factor. Since $\abs{2^j n + A_j} \leq N/2$, either $n= 0$ or $n= -1$ when $j = K$. The eigenvalue of $J_z$ is therefore unambiguously determined, i.e., for $j=K$:
\begin{equation}\label{eq:outputstate}
\ket{\psi_{j}}=\left\{
\begin{aligned}
&\ket{N, A_j}  \quad &A_j < 2^{j-1},\\
&\ket{N, A_j - 2^j} \quad &A_j > 2^{j-1}.
\end{aligned}
\right.
\end{equation}
As the standard deviation of $p(m_z)$ is $\sqrt{N}/2$, the equality in Eq.~(\ref{eq:outputstate}) approximately holds when $2^j \sim O(\sqrt{N})$. This means that in fact determining only the first 
$\ceil*{K/2}$ bits in Eq.~(\ref{eq:binary_rep}) can produce the target state with high fidelity, that is, the number of required ancilla qubit measurements can be further reduced in practice.

The controlled-$U^{2^{K-j}}$ gate is realized through the Hamiltonian in Eq.~(\ref{eq:system_hamiltonian}) below. The coupling strength $\gamma$ between the spins and the ancilla qubit determines how fast the gate is performed. Due to the exponentially decreasing rotation angles, the total evolution time $T$ of these controlled rotations is bounded, i.e.,
\begin{equation}\label{eq:action_angle}
\begin{split}
    T &= \sum_{j=1}^{K} t_j = \frac{\pi}{\gamma}(2-\frac{1}{2^K}) < \frac{2\pi}{\gamma} ,\\
    t_j &= \frac{\pi}{2^{j-1}\gamma}
    .
\end{split}
\end{equation}
Note that the preparation scheme requires initializing all qubits in $\ket{+}$ state, which can consume a considerable amount of time by itself, see Section \ref{sec:exp} for the experimental setup with NV electronic spins. 

An important comment on the use of standard phase estimation in Fig.~\ref{fig:pe_DickePreparation} is the following. Any gate will be implemented with some constant (small) error in practice, hence it is impossible to realize the rotation $U$ in Eq.~\eqref{eq:pe_unitary} when $K$ (and thus $N$) is too large. This error limits the maximum spin number $N$ that we can handle, as the rotation angle $2\pi/2^{K}$ scales as $O(1/N)$. For example, for $N=500$, we have $K=9$ and $2\pi/2^K \approx 0.012$, see also a further discussion in Section \ref{sec:control}.

One can also prepare a specific Dicke state by performing post-selection on the measurement outcomes, preparing $\ket{N,m_z\neq 0}$ in this way would require less operations than $\ket{N,m_z= 0}$ (see the details in Appendix~\ref{sec:specificDicke}). In addition, the idea of phase estimation can be used to prepare specific superpositions of Dicke states, which are potentially useful for metrology under noise \cite{ouyang2019robust} (see the details in Appendix~\ref{sec:DickeSuperposition}).

\section{System Hamiltonian and experimental realization}\label{sec:systemHamiltonian}

In this section, we will sketch an experimental realization using a superconducting flux qubit coupled to an ensemble of NV centers.

We consider a hybrid system where a set of $N$ two-level spins is collectively coupled to an ancilla qubit. To implement our scheme, we need the system Hamiltonian to be of the following form
\begin{equation}\label{eq:system_hamiltonian}
\begin{split}
    H =  H_0 &+ H_{\rm coupl},\\
    H_0 = \omega_0 J_z - \frac{1}{2} \omega Z,&\quad
    H_{\rm coupl} = \frac{\gamma}{2}  Z \otimes J_z
\end{split}
\end{equation}
with $J_z$ in Eq.~\eqref{eq:defJz}. Here, $\hbar = 1$ and $\gamma$ is the coupling strength between the ancilla qubit and the spins, $\omega$ is the angular frequency of the ancilla qubit. The spins in the ensemble are assumed to have the same energy splitting, denoted by angular frequency $\omega_0$. 

In this system, we assume the ability to (\romannumeral1) implement single-qubit rotations and projective measurements on the ancilla qubit, (\romannumeral2) implement global rotations of the spins (generated by $J_x, J_y$ and $J_z$), (\romannumeral3) initialize the ancilla qubit and the spins in $\ket{0}$.

The phase estimation scheme involves qubit-controlled rotations around $J_z$, which are realized through the interaction $H_I$. The evolution operator of $H_I$ is
\begin{align}\label{eq:evolution_HI}
    e^{-iH_It} = e^{-i\frac{\gamma}{2}t J_z}\left(\ket{0}\bra{0}\otimes I + \ket{1}\bra{1}\otimes e^{i \gamma t J_z} \right),
\end{align}
where the unconditional rotation $e^{-i\frac{\gamma}{2}t J_z}$ can be neglected. Since the free Hamiltonian $H_0$ commutes with $H_I$, we can also neglect the effect of $H_0$.

\subsection{Sketch of experimental implementation}
\label{sec:exp}

One possible experimental setup of the proposed protocol is an ensemble of NV centers coupled to a superconducting flux qubit. Each NV center hosts a single (electronic) $S=1$ spin. Sensing a magnetic field or spin by means of this electronic spin has been of high interest in the last decade, see e.g., Ref.~ \cite{casola2018probing, barry2020sensitivity} and references therein. Sensing using an ensemble of NV centers, without preparing them in a particular entangled state, has been used at ambient temperatures in, e.g., Ref.~\cite{Barry14133,kitazawa2017vector}. 

In addition, proposals exist to use the $^{13}C$ nuclear spins which surround a NV center to enhance the sensing performance \cite{vorobyov2020quantum,unden2016quantum}. Direct magnetic sensing using nuclear spins however would be inefficient, as their gyromagnetic ratio is about a factor 1000 less than that of the electronic spin.

The proposal in Ref.~\cite{marcos2010coupling} envisions coupling a flux qubit to NV center electronic spins for the transfer and storage of quantum information. This has been experimentally realized in Ref.~\cite{ZhuHybridNature2011}, where a flux qubit was coupled to $O(10^7)$ NV centers to resonantly transfer a flux qubit excitation to a collective spin excitation and back \footnote{In \cite{ZhuHybridNature2011} no external magnetic field was applied on the NV-electronic spins so that the states $\ket{m_z=\pm 1}$ are (nearly) degenerate and the excitation is to the level $\ket{m_z=\pm 1}$ and back.}.
In Ref.~\cite{hakoshima2020efficient} the preparation of Dicke states using a coupled flux qubit was considered for sensing, using this energy-transferring flip-flop interaction (of the form $\sigma_+ J_-+\sigma_- J_+$ where $\sigma_{\pm}$ acts on flux qubit and $J_{\pm}=\frac{1}{2}(J_x+i J_y)$ on the ensemble).
The basic idea for the Dicke state preparation in Ref.~\cite{hakoshima2020efficient} is then to repeat an excitation transfer from the flux qubit to the spins: (\romannumeral1) the flux qubit is first flipped to $\ket{1}$, (\romannumeral2) the hybrid system evolves for some chosen time during which the ancilla qubit goes back to $\ket{0}$ and the spins in $\ket{N,m_z = j}$ evolve to $\ket{N,m_z = j-1}$. Repeating this process $O(N)$ times, one obtains the state $\ket{N,m_z=0}$ from an arbitrary Dicke state, say the product state $\ket{N,m_z=N/2}=|00\ldots 0\rangle$.

In earlier work \cite{tanaka+:sensing}, the preparation of other sensing states, such as spin-cat and spin-squeezed states, was considered using a flux qubit coupled to a collection of NV-center electronic spins.

\subsubsection{ZZ Coupling between Flux Qubit and NV center Electronic Spins}

The coupling between the flux qubit and the NV center is magnetic, i.e. the two persistent current states of the flux qubit generate opposite magnetic fields which enter the Zeeman term in the NV center electronic spin Hamiltonian. As in Ref.~\cite{marcos2010coupling, ZhuHybridNature2011} one can imagine that the flux qubit is sitting on a diamond substrate with implanted NV centers, and say the loop of the flux qubit is about $1 \mu {\rm m} \times 1 \mu {\rm m}$. If the NV-centers are in a cubic volume $1 \times 10^{-18}\;{\rm m}^3$ below the loop, a NV-center density of $10^{21}\;{\rm m}^{-3}$ \cite{bar2013solid} would lead to already having about 1000 NV centers in this cube. 

The Hamiltonian of a general flux qubit itself is given by
\begin{equation}\label{eq:fluxHamiltonian}
    H_{\rm flux}=\frac{\lambda}{2} X_f-\frac{\epsilon}{2} Z_f,
\end{equation}
where the $Z$-basis is given by two persistent current states ($\ket{0}, \ket{1}$), --eigenbasis states of flux--, inducing opposite magnetic fields \cite{orlando, bylander2011noise,clarke2008superconducting}. Here we include a label $f$ to denote that these are Pauli operators on the flux qubit. The Pauli $X_f$ term is due to the kinetic charging energy.
The case $\epsilon=0$ corresponds to a symmetric double-well potential in flux. Since the required interaction in Eq.~\eqref{eq:system_hamiltonian} is $Z_f \otimes J_z$, we could envision that the current states are flux-qubit eigenstates. This implies an asymmetric double-well flux potential with $\epsilon > 0$ and $\epsilon \gg |\lambda|$ (requiring a large shunting capacitance). This is unlike some of the previous work mentioned above in which one works at $\epsilon=0$.  

Recent experiments demonstrate a long coherence time of the flux qubit at the flux sweet spot $\epsilon=0$. The energy relaxation time $T_1$ is about $40 \mu {\rm s}$ and the dephasing time $T_2$ is about 10 $\mu$s with dynamical decoupling \cite{yurtalan2020characterization}. Single-qubit gates with duration about $2\mathrm{ns}$ and fidelity about $99.92\%$ are also realized \cite{yurtalan2020characterization}. However, tuning a flux qubit away from $\epsilon=0$ decreases the dephasing time substantially. This is due to flux noise, i.e., the flux qubit becomes much more sensitive to fluctuations of $\epsilon$, which can be somewhat improved by dynamical decoupling \cite{bylander2011noise}. For this reason we discuss an alternative way of using the flux qubit at $\epsilon=0$ and the flip-flop interaction to realize a Dicke state preparation in Appendix~\ref{sec:adiabatic_JZ}. \\
 
There are four types of NV centers, each aligned with a different NV-axis of the carbon lattice (i.e. the direction from the vacancy to the nitrogen) \cite{kitazawa2017vector}, and one does not control the orientation of these NV-axes.
We choose one type of NV center and call its NV-axis the $z$-axis so that its associated electronic spin ($S = 1$) has Hamiltonian \cite{cramer2016quantum}
\begin{equation}
    H_{\rm NV}=\Delta S_z^2+W_z^{\rm ext} S_z,
    \label{eq:nv}
\end{equation}
where $W_z^{\rm ext}$ represents the effect of an externally applied magnetic field and $\Delta$ is the zero-field splitting ($\Delta \approx 2.88$ GHz). Here we neglect components of the magnetic field which are not aligned with the NV-axis.

With $W_z^{\rm ext}\neq 0$, the states $\ket{S_z=m=\pm1}$ are made non-degenerate and we imagine, as is fairly standard, using the lowest two energy eigenstates $\ket{S_z=m=0}$ and $\ket{S_z=m=-1}$ as the qubit. The externally applied magnetic field (O(100) Gauss) \cite{mohamed2021NV} which splits off the $m=\pm 1$ level should lie in the plane of the flux qubit loop, avoiding any stray effects on the flux qubit itself.

For a collection of $N$ NV centers, we thus restrict ourselves to the electronic $\{\ket{m=0}, \ket{m=-1}\}$ qubit subspace per NV center, and use the collective spin operators $J_x,J_y,J_z$ acting on these qubits. 

An additional magnetic field in the $y$-direction or $x$-direction, assuming it is uniformly experienced by all NV centers oriented along the $z$-axis, would induce additional Zeeman terms in the NV center Hamiltonian. This leads to global rotations, e.g. $\exp(-i \theta J_y)$, which we want to sense.  

By applying microwave ($O(1)$GHz) pulses with a frequency which is resonant with the NV-center electronic spins, rotations generated by the collective spin operators $J_x,J_y$ can be performed \cite{cramer2016quantum,bradley2019ten}.  To obtain the initial state $\ket{\psi_0}$ in Eq.~(\ref{eq:initial_distribution}), we first initialize all NV-center electronic spins in $\ket{0}$ through resonant optical excitations (the initialization duration is of the order of $O(100) \,\mu s$ \cite{Robledo_Nature2011} and is executed simultaneously for all NV-center electronic spins), then one performs the global rotation $e^{i\frac{\pi}{2}J_y}$ \cite{cramer2016quantum}.
In addition, the NV-center electronic spins can be collectively measured optically to measure $J_z$, but the limited photon collection efficiency limits the readout contrast \cite{taylor2008high,xu2019high, barry2020sensitivity}.

The Hamiltonian of a single NV-center and a flux qubit is
\begin{equation}
    H= H_{\rm flux}+H_{\rm NV}+H_{\rm coupl}.
\end{equation}
The coupling term $H_{\rm coupl}$ models the NV-electronic spin experiencing a magnetic field due to the different persistent current flux qubit states: it can be written in the form
\begin{equation}
    H_{\rm coupl}=-\gamma_e  \vec{B}_{\rm flux} \cdot \vec{S}
\end{equation}
with spin $S=1$ operators $\vec{S}=(S_x, S_y, S_z)$ and gyromagnetic ratio $\gamma_e$ ($\sim 2.8{\rm MHz}/{\rm Gauss}$). Let's call the axis perpendicular to the flux qubit loop $\hat{r}$, so that $\vec{B}_{\rm flux} \approx B\hat{r}Z_f$, where $Z_f$ is the flux qubit Pauli $Z$ operator and $B$ is the magnetic field strength at the NV center. Here we assume that the NV centers are centrally placed below the flux qubit, so that magnetic field components in directions other than $\hat{r}$ are negligible.

Since the coupling is much weaker than the electronic spin qubit frequency, we neglect the change of the precession axis induced by this coupling term. The coupling is thus approximated as
\begin{equation}
    H_{\rm coupl} \approx \frac{\gamma}{2} Z_f\otimes S_z.
 \end{equation}
 Ref.~\cite{marcos2010coupling} estimates that the coupling strength can be about $ 12 \mathrm{ kHz}$, depending on the strength of the magnetic field $B$ and the proximity to the NV-center.
We assume that we can use an orientation $\hat{r}$, so that the (projected) coupling strength $\gamma/2 = -\gamma_e B r_z $ is also of the order of $O(10)\mathrm{kHz}$. 
 
\begin{figure}
	\[
	\Qcircuit @C=0.6em @R=0.6em{
    &&&&&&\lstick{\text{electronic spins}} & \qw {/} &\qw & \gate{e^{i\frac{\gamma t}{2} J_z}} & \gate{e^{i\pi J_y}} & \gate{e^{i \frac{\gamma t}{2} J_z}} & \gate{e^{i\pi J_y}} 
  \\
  &&&&&&\lstick{\text{flux qubit} } & \qw & \qw & \ctrl{-1}  &\gate{X} &\ctrl{-1} &\gate{X} 
	}
	\]
	\caption{The controlled-$e^{i\gamma t J_z}$ gate with integrated dynamical decoupling, up to the unconditional rotation $e^{-i\frac{\gamma t}{2}J_z}$. Echo pulses are simultaneously applied to the NV-center electronic spins and the flux qubit. The pulses are at a frequency resonant with those NV-center electronic spins which should remain coupled to the flux qubit, while the coupling to the other NV-centers is echoed away. }
	\label{fig:controlledrotation_echo}
\end{figure}
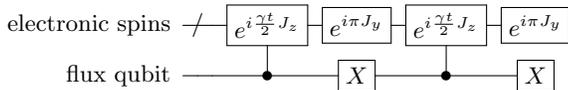

The four types of NV centers are simultaneously coupled to the flux qubit, each having a different coupling strength as their NV-axes are different and having a different resonance frequency \cite{kitazawa2017vector}. In principle all types of NV centers could be used for sensing different components of the magnetic field \cite{Barry14133}. However, since we have only a single controlling flux-qubit to create an entangled state, it is preferred to dynamically decouple the interaction with the other NV centers away.

For example, to cancel the coupling to 3 of the 4 NV center types, one could perform phase estimation with integrated dynamical decoupling. The circuit in Fig.~\ref{fig:controlledrotation_echo} realizes the controlled-$e^{i\gamma t J_z}$ gate up to the unconditional rotation $e^{-i\frac{\gamma t}{2} J_z}$. The echo pulse $e^{i\pi J_y}$ is implemented using resonant microwaves with NV centers whose NV axis is the $z$-axis.  These other NV centers are thus decoupled from the flux qubit.

The integrated echo pulses also provide resilience to the dephasing of the flux qubit and NV-center electronic spins. Note that we can split the controlled rotations to controlled-$e^{i\gamma t J_z/n}$ with $n=4, 6, 8, \ldots$, so that we obtain a further suppression of the dephasing. For simplicity, we will consider $n=2$ in numerics in Section \ref{sec:noisyPreparation}.

The coherence time of NV-center electronic spins is not a limiting factor for realizing our preparation scheme. The energy relaxation time $T_1$ of NV-center electronic spins exceeds 8 hours at $25 \mathrm{mK}$ \cite{astner2018solid}.  For a NV ensemble with a NV-density about
$10^{21}$ $\mathrm{m}^{-3}$, the dephasing time $T_2$  (with dynamical decoupling) can be about 50$\mathrm{ms}$ at 77 $\mathrm{K}$ \cite{bar2013solid}. Because the dephasing of a NV electronic spin mainly comes from its surrounding spin bath environment \cite{de2010universal}, we may expect a longer dephasing time at the operating temperature of the flux qubit (tens of $\mathrm{mK}$). 

The weak point of this sketched proposal is the strength of the coupling $\gamma$ versus the (short) dephasing time of the flux qubit $T_2 < O(1) \mu$s if it is operated away from its flux sweet spot. Even though the flux qubit only needs to stay coherent during each round of phase estimation individually, i.e. during a circuit as in Fig.~\ref{fig:pe_DickePreparation}, a $O(10)$ kHz coupling $\gamma$ requires an interaction time much longer than the flux qubit coherence time in particular for small $j$.

As an alternative, it may be possible to use the three levels ($S=1$) of the NV-center electronic spins 
to apply controlled rotations adiabatically 
while operating the flux qubit in a more phase-coherent regime with $T_2=O(10)\mu$s. In this scenario we work at $\epsilon=0$ for the flux qubit in Eq.~\eqref{eq:fluxHamiltonian} and adiabatically change the flux qubit frequency through flux-control while staying at $\epsilon=0$. Remember that the states $\ket{m=0}_{\rm NV}$ and $\ket{m=-1}_{\rm NV}$ form the NV-center qubit subspace and $\ket{m=+1}_{\rm NV}$ is a third level.

If we apply a Hadamard transformation to make the flux qubit Hamiltonian diagonal in $Z_f$, the coupling term will read $H_{\rm coupl}=-\gamma_e B X_f\otimes \vec{S}\cdot \hat{r}$. Neglecting non-secular terms, one is left with an interaction which removes a flux-qubit excitation while exciting the NV center electronic state $m=0$ to $m=\pm 1$ and vice-versa.
We imagine adiabatically flux-tuning the flux qubit frequency to the avoided crossing between $\ket{1}_{\rm flux}\otimes \ket{m=0}_{\rm NV}$ and $\ket{0}_{\rm flux}\otimes \ket{m=+1}_{\rm NV}$ and back so as to get an effective $ZZ$-interaction in the $\ket{m=0}_{\rm NV}$ and $\ket{m=-1}_{\rm NV}$ and flux-qubit subspace. This way of obtaining a $ZZ$-interaction using a third level is commonly done for superconducting transmon qubits \cite{Martinis2014-zw,rol:ZZ}. Here we would need to generate this interaction between a single ancilla qubit and $N$ NV electonic qubits, each with a third level. 
Note that this idea is different from flux-tuning the frequency of the flux qubit to be resonant with the NV-center electronic spin qubit frequency to activate the flip-flop interaction \cite{hakoshima2020efficient, ZhuHybridNature2011}. We discuss the details about adiabatically applying controlled rotations in Appendix~\ref{sec:adiabatic_JZ}.

In this alternative scenario, one is also limited by the strength of the magnetic coupling. The coupling can only be enhanced by increasing the proximity of the NV-centers to the flux-qubit loop and having a higher current associated with the flux qubit states (leading to a stronger magnetic field), but the Josephson critical current density puts limits on this.  

\begin{table}[htb]
	\centering
	\begin{tabular}{|l||l|}
		\hline
		 &Typical value \\
		\hline
		NV electronic spin $T_1$ &$\quad > 1 \mathrm{h}$ \\
		\hline
		NV electronic spin $T_2$ &$\quad >O(50)\, \mathrm{ms}$\\
		\hline
		NV electronic spin initialization time  &$\quad O(100)\, \mu s$ \\
		\hline
		Flux qubit $T_1$ & $\quad O(50)\, \mu s$ \\
		\hline
		Flux qubit $T_2$ & $\quad < O(1)\, \mu s$\\
		\hline
		Flux qubit single-qubit gate time  &$\quad O(1)\,  ns$ \\
		\hline
		Magnetic coupling $\gamma$ &$\quad O(10)\, \mathrm{kHz}$ \\
		\hline
	\end{tabular}
	\caption{Parameters that are relevant for realizing the preparation scheme using the sketched experimental setup, where an ensemble of NV electronic spins is collectively coupled to a single superconducting flux qubit. The main challenge is the weak magnetic coupling $\gamma$ versus the short dephasing time $T_2$ of the flux qubit.
	}
	\label{tab:limiting_factors}
\end{table}

For realizing the preparation scheme using the sketched experimental setup, relevant parameters with their typical values are listed in Table~\ref{tab:limiting_factors}.

\section{Preparation with noise}
\label{sec:noisyPreparation}
In this section, we look at the performance of the phase estimation scheme for stronger coupling strength $\gamma$ than what has been stated in the previous section, and some limited decoherence of the flux qubit. The spins in the ensemble are assumed to be perfect, as their coherence time is not a limiting factor for our scheme.

\subsection{Limited coherence time of the flux qubit}

The flux qubit has energy relaxation time $T_1$ and limited pure dephasing time $T_{\phi}$ with $\frac{1}{T_2} = \frac{1}{2T_1} + \frac{1}{T_{\phi}}$. A simple model for the effect of $T_{\phi}$ is that of a phase flip channel. That is, during the controlled-$e^{i\gamma t J_z}$ gate, the flux qubit obtains a Pauli $Z$ error with an error rate \cite{nielsen2002quantum} 
\begin{equation}\label{eq:dephasing_readout}
    P_{T_{\phi}}(t) = \frac{1 - e^{-t/T_{\phi}} }{2}.
\end{equation}
Such Pauli $Z$ error can flip the ancilla qubit readout in the phase estimation circuit. Fortunately, we can suppress this error to some extent by repeating each circuit in Fig.~\ref{fig:pe_DickePreparation} and taking a majority vote of the answers.

Flux qubit decay to zero temperature with rate $\kappa = 1/T_1$ can be described by a Lindblad master equation
\begin{equation}
    \frac{\mathrm{d}\rho}{\mathrm{d}t} = -i [H, \rho] + \kappa \mathcal{D}[\sigma_{-}\otimes I]\rho.
\end{equation}
Here $\sigma_{-} = \ket{0}\bra{1}$ is the annihilation operator on the ancilla qubit, $\rho$ is the density matrix of the total system, and $\mathcal{D}[c]$ is defined as $\mathcal{D}[c]\rho = c\rho c^{\dagger} - \frac{1}{2}\{c^{\dagger}c, \rho\}$. The Kraus operators for a short time $dt$ are
\begin{equation}
\begin{split}
    \delta M_0 = \ket{0}\bra{0}&\otimes I + e^{-\frac{1}{2}\kappa\mathrm{d}t} \ket{1}\bra{1}\otimes  e^{i\gamma \mathrm{d}t J_z},\\
    \delta M_1 &= \sqrt{\kappa\mathrm{d}t}\ket{0}\bra{1}\otimes I.
\end{split}
\end{equation}
Here the free evolution $e^{-iH_0 \mathrm{d}t}$ and the unconditional rotation $e^{-i\frac{\gamma}{2}t J_z}$ in Eq.~(\ref{eq:evolution_HI}) are omitted.
Note that the Kraus operator $\delta M_1$ does not commute with $H_I$.

For a finite evolution time $t$, the action of the controlled-$e^{i\gamma t J_z}$ gate is given by a continuous set of Kraus operators\begin{equation}\label{eq:controlledRotation_Kraus}\begin{split}    M_0(t) &= \ket{0}\bra{0}\otimes I + \ket{1}\bra{1}\otimes e^{-\frac{1}{2}\kappa t} e^{i\gamma t J_z},\\    M_1(t') &= \sqrt{\kappa e^{-\kappa t'}}\ket{0}\bra{1}\otimes e^{i\gamma t' J_z} \quad \text{for }  t'< t.\end{split}\end{equation}
The controlled rotation in the presence of ancilla qubit decay is then described by the quantum channel\begin{equation}
\rho_{out} = M_0(t)\rho_{in} M_0^{\dagger}(t) + \int_{0}^{t} \mathrm{d}t' M_1(t')\rho_{in} M_1^{\dagger}(t').\end{equation}
where $\rho_{in}$ and $\rho_{out}$ are the input and output states of the controlled gate.

Now we consider implementing the controlled-$e^{i\gamma t J_z}$ gate with integrated echo pulses as in Fig.~\ref{fig:controlledrotation_echo}. If the ancilla qubit does not decay, the circuit applies the Kraus operator
\begin{equation}
\begin{split}
    (X\otimes \,&e^{i\pi J_y})\, M_0(t/2)\,  (X\otimes e^{i\pi J_y})\, M_0(t/2) 
    \\&=  \sqrt{e^{-\frac{1}{2}\kappa t}} e^{-i\frac{\gamma t}{2} J_z} \left(\ket{0}\bra{0}\otimes I + \ket{1}\bra{1}\otimes  e^{i\gamma t J_z} \right).
\end{split}
\end{equation}
This is the desired gate up to the unconditional rotation $e^{-i\frac{\gamma t}{2} J_z}$, which happens with the probability $e^{-\frac{1}{2}\kappa t}$. Otherwise, the ancilla qubit decays and an unconditional rotation around $J_z$ is applied to the spins. In phase estimation, no projector is implemented and the ancilla qubit readout gives the outcome $0, 1$ with  equal probability.

The flux qubit decay probability during the controlled-$e^{i\gamma t J_z}$ gate (with integrated echo pulses) is given by
\begin{equation}
    P_{T_1}(t) = 1 - e^{-\frac{t}{2T_1}}.
\end{equation}

\subsection{Inaccurate control of the flux qubit}
\label{sec:control}

Each time we perform the controlled-$e^{i\gamma t J_z}$ gate, there may be a random small deviation $\delta t$ of the evolution time $t$ which becomes important when $t$ is small. This means we actually perform the controlled-$\beta e^{i\gamma t J_z}$ gate with $\beta = e^{i\delta \theta J_z}$ and $\delta \theta = \gamma \delta t$.  

Preparing an $N$-spin Dicke state
means determining the eigenvalue of the unitary rotation $U$ in Eq.~(\ref{eq:pe_unitary}), whose rotation angle $2\pi/2^{K}$ scales as $O(1/N)$. As we have discussed, determining the first $\ceil*{K/2}$ bits in Eq.~(\ref{eq:binary_rep}) produces the target state with high fidelity. That is, we only need to determine the eigenvalue of a unitary rotation whose rotation angle scales as $O(1/\sqrt{N})$. This scaling characterizes how much timing inaccuracy our scheme can tolerate and thus how large $N$ can be.

\begin{figure}
\centering
\includegraphics[width=\linewidth]{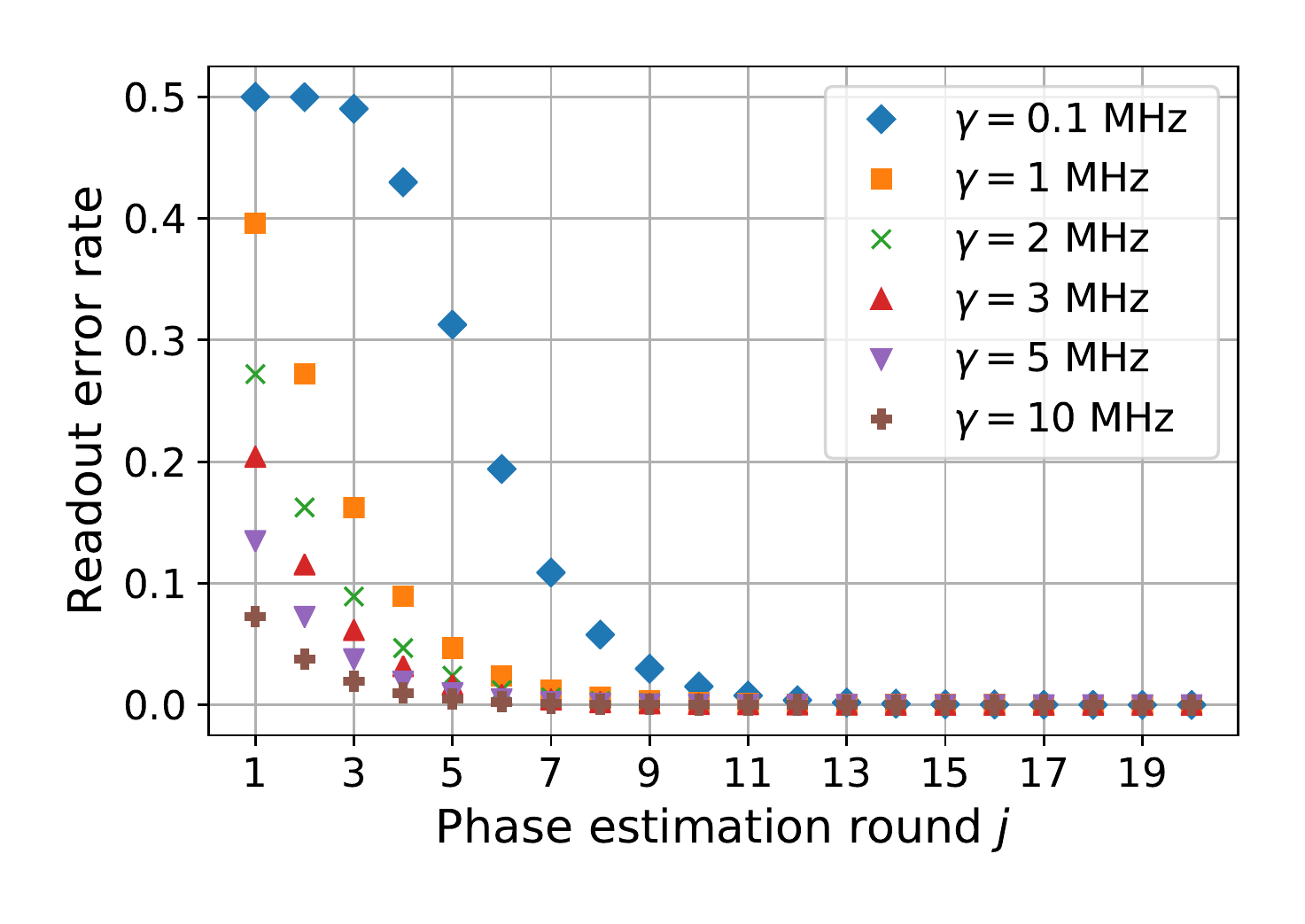}
\caption{(Color online) The flux qubit readout error rate induced by flux qubit dephasing. Here we fix the pure dephasing time of the flux qubit as $T_{\phi} = 2 \mu\mathrm{s}$. In the $j$th round phase estimation, the evolution time is $t_j = \pi 2^{1-j}/\gamma$ which leads to the readout error rate $ P_{T_{\phi}}(t_j)$ in Eq.~(\ref{eq:dephasing_readout}).  
}
\label{fig:readouterror}
\end{figure}

\subsection{Numerical Simulations}

In the presence of inaccurate control, ancilla qubit dephasing, or ancilla qubit decay, the spins in the ensemble stay in the subspace which is symmetric under spin permutations.
They can be treated as a large pseudo-spin of size $J = N/2$. We therefore limit ourselves to a state vector in a $N+1$-dimensional space rather than the full size $2^N$. The simulation is based on the QuTip Python package \cite{johansson2012qutip, johansson2013qutip2}, and the code can be found in \cite{YWGithub2020}.

The preparation starts from the product state $\ket{\psi_0}$ in Eq.~(\ref{eq:initial_distribution}), and determines the eigenvalue $m_z$ of $J_z$ using standard phase estimation. Each round of phase estimation is repeated multiple times, and a simple majority vote is performed.
The fidelity of the prepared state, with respect to the predicted state $\ket{N,m_z}$, is used as a measure of how good the preparation is. We compute $F=\bra{N, m_z} \rho \ket{N, m_z}$ where $\rho$ is the density matrix prepared by the noisy, imperfect protocol. 

Here we assume that $T_1 = 50$ $\mu$s and $T_{\phi} = 2$ $\mu$s for a flux qubit far away from flux-sweet spot.
To ensure $P_{T_{\phi}}$ is reasonably small in the first few rounds of phase estimation, we need the coupling strength to be a few $\mathrm{MHz}$, say, $\gamma = 5 \mathrm{MHz}$, as shown in Fig.~\ref{fig:readouterror}. The corresponding flux qubit decay probability $P_{T_1}(t)$ would then only be about $0.6\%$ in the first (longest) round of phase estimation. Hence the effect of pure dephasing $T_{\phi}$ dominates.

Suppose each round of phase estimation is repeated $M$ times with majority voting. We say that the $j$th round phase estimation succeeds when: (\romannumeral1) There is at least one measurement, during which the flux qubit does not decay, i.e., the projector $\mathrm{P}(m_j)$ in Eq.~(\ref{eq:idealprojector}) is implemented at least once.
(\romannumeral2) Majority voting of the $M$ measurement outcomes yields the correct answer $m_j$.  

\begin{figure}
\centering
\includegraphics[width=\linewidth]{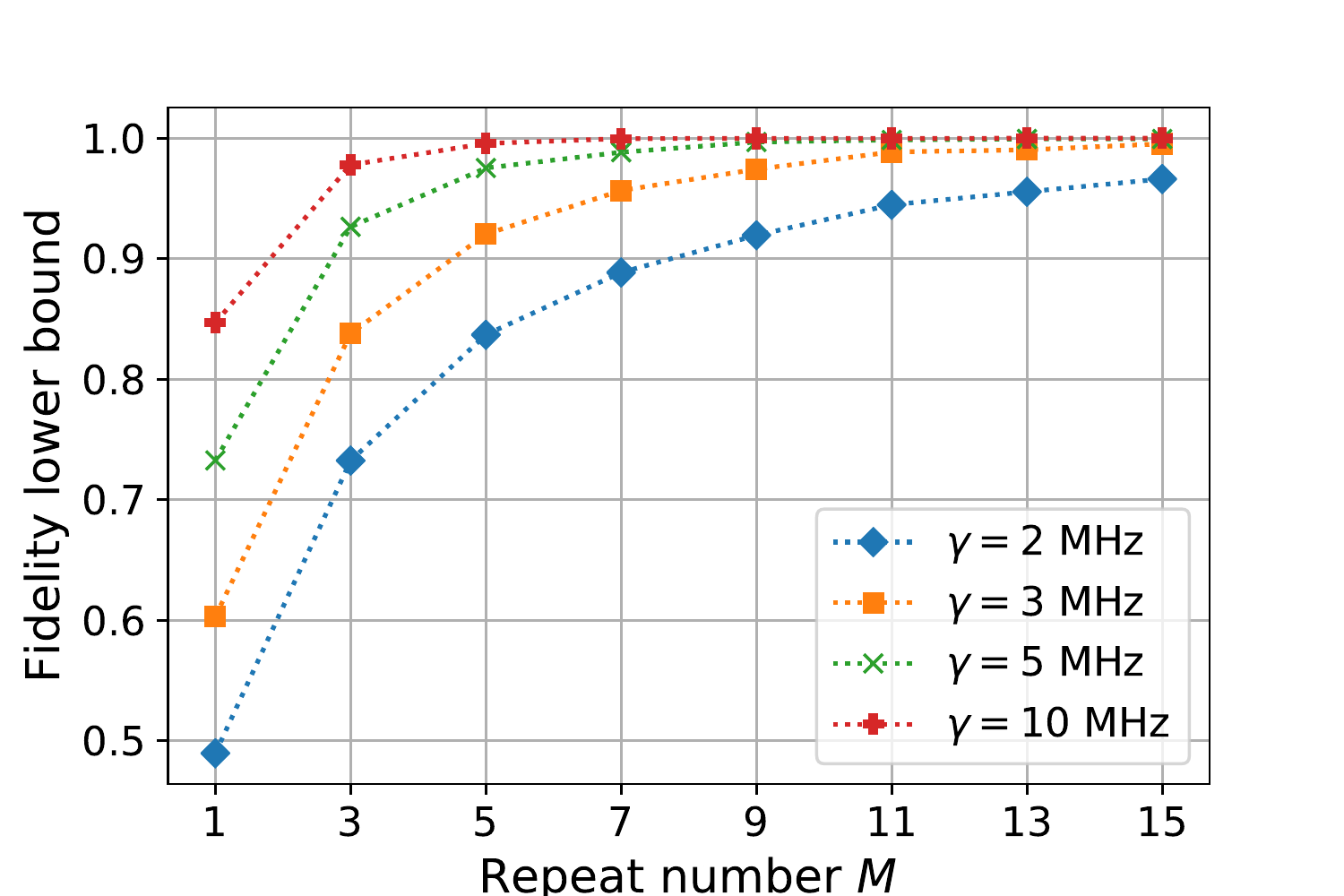}
\caption{(Color online) Lower bound on the output fidelity for the preparation scheme with limited flux qubit coherence time but strong coupling. There are $K=20$ rounds of phase estimation, each is repeated $M$ times with majority voting.
The fidelity lower bound $\mathrm{P}$ is the probability that all rounds of phase estimation succeed, so that the prepared state has fidelity 100\%. Here we set $T_1 = 50 \mu\mathrm{s}$ and $T_{\phi} = 2\mu\mathrm{s}$.
}
\label{fig:preparation_flux_decoherence}
\end{figure}

Instead of a full simulation, we numerically calculate the probability that all rounds of phase estimation succeed, i.e., the prepared state has fidelity 100\%. This probability is calculated as $\mathrm{P} = \prod_{j=1}^{K} P_j$, with $P_j$ the success rate of the $j$th round phase estimation. Clearly, the probability $\mathrm{P}$ sets a lower bound of the output fidelity $F$. The lower bound $\mathrm{P}$ is plotted in Fig.~\ref{fig:preparation_flux_decoherence}, where we set $K=20$ and use a coupling strength $\gamma$ much beyond current estimates. Note that $20$ rounds of phase estimation correspond to about $10^6$ spins. We find that $\mathrm{P}$ quickly approaches unity as the repeat number $M$ grows, basically because $P_{T_{\phi}}(t_j)$ and $P_{T_1}(t_j)$ both decrease exponentially as $j$ grows.

To model inaccurate timing control of the flux qubit, we run a pure state simulation. Each time we apply a controlled-rotation in the preparation, there is a randomly sampled time deviation $\delta t$. We assume that $\delta t$ is distributed according to the normal distribution $\mathcal{N}(0, \sigma^2)$. Considering that single-qubit rotation on a flux qubit has a duration about $2 \mathrm{ns}$ \cite{yurtalan2020characterization}, we set $\sigma = 0.5,1,3,6,10 \mathrm{ns}$. Additionally, we fix $\gamma = 5 \mathrm{ MHz}$ and $N = 500$ spins. 
In the simulation, we perform 6 rounds of phase estimation, and each round is repeated $M$ times with majority voting (here $T_{\phi}, T_1=\infty$).  Note that in the noiseless case, 6 rounds of phase estimation gives an output fidelity $F\approx 99.65\%$ for $N = 500$.

\begin{figure}
\centering
\includegraphics[width=\linewidth]{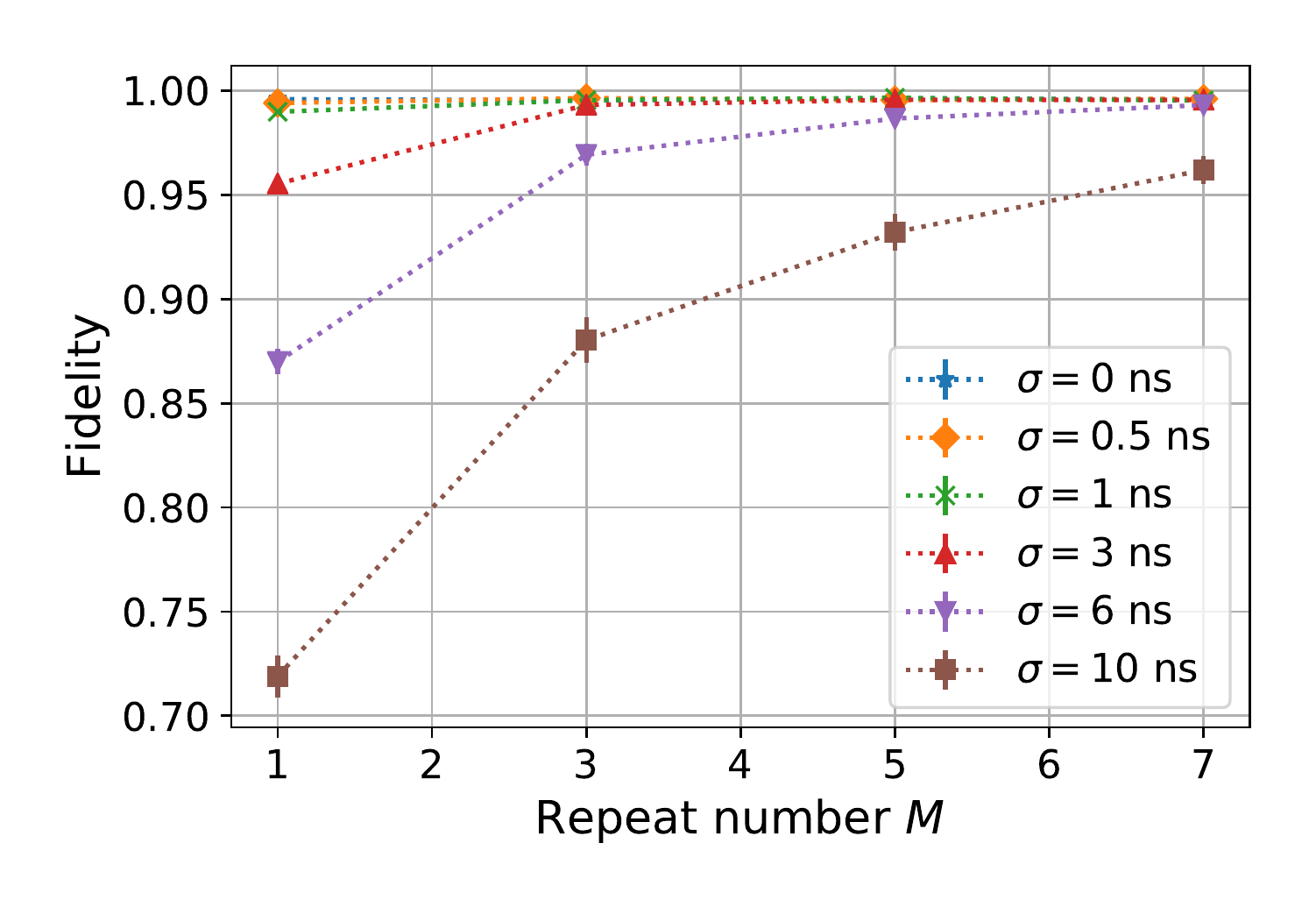}
\caption{(Color online) The output fidelity of the preparation scheme with inaccurate control of the flux qubit. Here we fix the spin number $N=500$, and the coupling strength $\gamma = 5 \mathrm{ MHz}$. There are only 6 rounds of phase estimation, each is repeated $M$ times with majority voting. When a controlled-rotation is applied, there is a time deviation $\delta t$, which is sampled according to normal distribution $\mathcal{N}(0, \sigma^2)$.
The error bar is the 95\% confidence interval.
}
\label{fig:fidelity_inaccurateControl}
\end{figure}

As shown in Fig.~\ref{fig:fidelity_inaccurateControl}, our scheme is resilient to inaccurate flux qubit control. Even with $\sigma = 10 \mathrm{ns}$ and $\gamma = 5 \mathrm{ MHz}$, the output fidelity still surpasses 90\% when we repeat each projective measurement only $M = 5$ times. Considering that $\gamma = 5 \mathrm{ MHz}$ is much larger than an estimated $ 12 \mathrm{ kHz}$, the expected effect of inaccurate timing in flux qubit control would thus be much smaller in practice. Hence, we would not expect this timing inaccuracy to be a main experimental challenge in the near future.

\section{Conclusion}\label{sec:Discussion}
To summarize, we have presented the idea of using phase estimation to prepare highly entangled Dicke states. Phase estimation can be realized through a global control, and
only requires $O(\log_2 N)$ ancilla qubit measurements. 
Dicke states $\ket{N,m_z}$ with $m_z\sim O(1)$ are especially interesting for metrology as they can give Heisenberg-limited sensitivity via global control. Phase estimation can prepare such Dicke states with a probability $O(1/\sqrt{N})$, implying the need for $O(\sqrt{N})$ attempts on average. 

With numerical simulations, we show that our scheme has some robustness to noise on the ancilla qubit. However, our scheme is still demanding for a spin ensemble coupled to a flux qubit as it requires a larger coupling strength or a longer flux qubit coherence time than what seems currently feasible.

One aspect of our analysis is that we assume a uniform coupling strength $\gamma$ of the ancilla qubit to the spin ensemble, while in practice not all NV centers will be equidistant. 
Thus each spin in the ensemble will have a slightly different coupling strength $\gamma_i=\gamma + \delta \gamma_i$. The deviation $\delta \gamma_i$ results in local over-rotations leading to $U=\exp(i 2 \pi H_z/2^K)$ instead of Eq.~\eqref{eq:pe_unitary}, where $H_z=J_z+\frac{1}{2}\sum_i \frac{\delta \gamma_i}{\gamma} Z_i$. Phase estimation will estimate the eigenvalues of $H_z$ to a precision set by $K$ and approximately project the state onto an eigenstate of $H_z$. Product states $|x\rangle$ with the same Hamming weight are no longer degenerate with respect to $H_z$, which implies that a perfect eigenstate projection would lead to preparing a product state. For small enough $N$, when this eigenvalue-breaking contribution is small, i.e. $\sum_{i=1}^N| \frac{\delta \gamma_i}{2\gamma}| \ll 1$, one may expect that the projected eigenstates of $H_z$, starting from the state $\ket{\psi_0}$ in Eq.~\eqref{eq:initial_distribution}, are still (weighted) superpositions of bitstrings and thus entangled. In addition, imperfect state initialization can break permutation symmetry. It may be of interest to numerically simulate the sensing ability of the states that one projects onto using $H_z$. However, such numerical simulations are much more challenging as we have to consider matrices of size $2^N \times 2^N$. 

To prepare $\ket{N,m_z=0}$ more efficiently, we can combine our scheme with the proposal in Ref.~\cite{hakoshima2020efficient} (see Sec \ref{sec:exp}), as both can be realized using the same experimental setup, i.e., a spin ensemble (e.g., diamond NV centers) coupled to a superconducting flux qubit. The difference is that the proposal in Ref.~\cite{hakoshima2020efficient} requires the flux qubit to operate at $\epsilon=0$, leading to the flip-flop interaction \cite{marcos2010coupling,ZhuHybridNature2011}. Note that our phase estimation scheme (starting from the product state) prepares a Dicke state $\ket{N,m_z < O(\sqrt{N})}$ with a probability approaching unity. If we could tune the flux qubit to the $\epsilon=0$ point after phase estimation (or use the scheme in Appendix \ref{sec:adiabatic_JZ}), then we can perform the scheme in Ref.~\cite{hakoshima2020efficient} to obtain $\ket{N,m_z=0}$, which uses another $O(\sqrt{N})$ flux qubit flips.

\section*{Acknowledgements}
This work is part of the project QCDA (with project number 680.91.033) of the research programme QuantERA which is (partly) financed by the Dutch Research Council (NWO). We would like to thank Lingling Lao, Mohamed Abobeih, Daniel Weigand, Alessandro Ciani and Boris Varbanov for useful discussions. 

\appendix

\section{Optimizing the preparation of a specific Dicke state}\label{sec:specificDicke}

Standard phase estimation can also prepare a specific Dicke state $\ket{N, m_z = m}$ when post-selection is used. To maximize the success rate, we rotate each spin initialized in $\ket{+}$ around the $Y$-axis by an angle $2\chi$. 
The value of $\chi$ is chosen so that $p = \frac{1}{2}\left[\cos(\chi) - \sin(\chi)\right]^2 = \frac{m + N/2}{N}=\frac{m}{N}+\frac{1}{2}$. 
The preparation of $\ket{N, m_z = m}$ then starts from
\begin{equation}\label{eq:rotated_intial_superposition}
    \begin{split}
      e^{-i2\chi J_y} \ket{+}^{\otimes N}
    & =  ( \sqrt{p}\ket{0} + \sqrt{1-p}\ket{1})^{\otimes N}\\
    &= \sum_{m_z = -N/2}^{N/2} \sqrt{ \tilde{P}(m_z)  }\ket{N, m_z},\\
    \tilde{P}(m_z) &= \binom{N}{m_z+N/2}p^{m_z+N/2}(1-p)^{N/2-m_z}.
    \end{split}
\end{equation}
Here, the distribution $\tilde{P}(m_z)$ reaches its maximum at $m_z = m$, as it is most likely that we draw $p N=m+\frac{N}{2}$ $1$s in this Bernoulli process, corresponding to $m_z=m$. Standard deviation of the distribution is $\sqrt{\frac{N^2 - 4m^2}{4N}}$. Note that $\tilde{P}(m)$ upper bounds the probability of obtaining $\ket{N,m}$.

The target state is obtained by measuring the operator $e^{i\pi 2^{1-l} (J_z - m)}$ with $l$ integer via phase estimation as before i.e.,
\begin{equation}\label{eq:specificDicke}
   \ket{N, m} = \frac{1}{\sqrt{\tilde{P}(m) }}  \prod_{l=1}^{K}\, \frac{1 + e^{i\pi 2^{1-l} (J_z - m)}}{2} e^{-i2\chi J_y} \ket{+}^{\otimes N}.
\end{equation}
The $K = \ceil*{\log_2 N} + 1$  measurements are realized through the circuit in Fig.~\ref{fig:pe_DickePreparation}. The equality in Eq.~(\ref{eq:specificDicke}) approximately holds when the number of measurements $K$ satisfies that $2^K \sim O(\sqrt{\frac{N^2 - 4m^2}{4N}})$. 

\section{Quantum code: superposition of Dicke states}\label{sec:DickeSuperposition}

In Ref.~\cite{ouyang2014permutation, ouyangPISecond}, a so-called permutation-invariant code is proposed for quantum error correction. The logical code words of this code are specific superpositions of Dicke states, namely
\begin{equation}\label{eq:PI_sensorstate}
\begin{split}
    \ket{0_L}=\frac{1}{\sqrt{2^{n-1}}} \sum_{\substack{0 \le j \le n\\j \text{ even}}}\sqrt{\binom{n}{j}}\ket{N = gnu, m_z = gj-\frac{N}{2}}.\\
    \ket{1_L}=\frac{1}{\sqrt{2^{n-1}}}\sum_{\substack{0 \le j \le n\\j \text{ odd}}}\sqrt{\binom{n}{j}}\ket{N = gnu, m_z = gj-\frac{N}{2}}.
\end{split}
\end{equation}
The code can correct arbitrary $t$-qubit Pauli errors with $g,n > 2t+1$, where $g,n$ are both integers. The rational number $u\geq 1$ is a scaling parameter which controls the total qubit number $N = gnu$ \cite{ouyang2014permutation, ouyangPISecond}. Note that the construction of such permutation-invariant codes has been generalized in Ref.\cite{Ouyang2016cl}, where the logical states are encoded into multiple qudits.

In Ref.~\cite{ouyang2019robust}, it was suggested to use the logical state $\ket{+_L}=\frac{\ket{0_L} + \ket{1_L}}{\sqrt{2}}$ as a probe state in metrology. It is claimed that the suggested probe state can give Heisenberg-limited sensitivity, even in the presence of a non-trivial number of errors \cite{ouyang2019robust}. For simplicity, we write $\ket{+_L}$ with parameters $g,n,u$ as
\begin{equation}
\begin{split}
    \ket{\varphi_{g,n,u}}= \frac{1}{\sqrt{2^n}} \sum_{0 \le j \le n}\sqrt{\binom{n}{j}}\ket{N = gnu, m_z = gj-\frac{N}{2}}.
\end{split}
\end{equation}
Here we show how this probe state $\ket{\varphi_{g,n,u}}$ can be prepared using phase estimation. Note that the preparation of this code has been studied in Ref.~\cite{wu2019initializing, johnsson2019geometric, Ouyang2021-af}.

We first look at the simplest 9-qubit code with $g = n = 3$ and $u=1$, which corrects an arbitrary single-qubit Pauli error. The corresponding probe state is
\begin{equation}\label{eq:PI_9qubitsensorstate}
    \ket{\varphi_{3,3,1}}= \frac{\ket{9, -\frac{9}{2}} + \sqrt{3}\ket{9, -\frac{3}{2}} + \sqrt{3}\ket{9, \frac{3}{2}} + \ket{9, \frac{9}{2}}}{\sqrt{8}}.
\end{equation}
One can find that $\ket{\varphi_{3,3,1}}$ is an eigenstate of the operator $e^{i\frac{2\pi}{3} J_z}$ with eigenvalue $-1$. The basic idea is to project a permutation-invariant state into an $-1$ eigenstate of $e^{i\frac{2\pi}{3} J_z}$, i.e., a superposition of $\ket{N = 9, m_z}$ with $m_z = \pm \frac{3}{2}, \pm \frac{9}{2}$.  
That is to say, we can prepare the target state by determining the eigenvalue of the unitary operator $e^{i\frac{2\pi}{3} J_z}$. This can be realized by measuring the operator multiple times with post-selection, which can effectively project out Dicke states $\ket{N,m_z}$ with $m_z \neq \pm \frac{3}{2}, \pm \frac{9}{2}$. 

However, the obtained superposition is not necessarily the target state $\ket{\varphi_{3,3,1}}$,  as the amplitudes are carefully chosen. To fix this problem, we will start from the initial state $\ket{\overline{\varphi}_{3,3,1}}$ which satisfies
\begin{equation}\label{eq:desired_initial}
\begin{split}
    \braket{9,\frac{9}{2}|\overline{\varphi}_{3,3,1}} = \braket{9,-\frac{9}{2}|\overline{\varphi}_{3,3,1}}&, \,\, \braket{9,\frac{3}{2}|\overline{\varphi}_{3,3,1}} = \braket{9,-\frac{3}{2}|\overline{\varphi}_{3,3,1}},\\
    \braket{9,\frac{3}{2}|\overline{\varphi}_{3,3,1}} &= \sqrt{3} \braket{9,\frac{9}{2}|\overline{\varphi}_{3,3,1}}.
\end{split}
\end{equation}
For a superposition of Dicke states, we observe that a rotation $e^{i\theta J_y}$ changes the distribution of $m_z$, which can be seen from Eq.~(\ref{eq:rotated_intial_superposition}). The desired initial state $\ket{\overline{\varphi}_{3,3,1}}$ can be approximately constructed as 
\begin{equation}\label{eq:PI-9QUBIT}
    \ket{\overline{\varphi}_{3,3,1}} = \frac{1}{\sqrt{\mathcal{N}}} \frac{e^{-i\theta J_y} + e^{i\theta J_y}}{2} \ket{+}^{\otimes 9}, \quad \theta = 0.57056
\end{equation}
with $\mathcal{N}$ the normalization factor.
Here the value of $\theta$ is obtained numerically. The application of the operator $(e^{-i\theta J_y} + e^{i\theta J_y})/2$ can be realized by measuring $e^{i2\theta J_y}$ with post-selection, followed by the unitary rotation $e^{-i\theta J_y}$. The target state $\ket{\varphi_{3,3,1}}$ is then approximately obtained: 
\begin{align}\label{eq:PI_9QUBITProjection}
    \ket{\varphi_{3,3,1}} \approx \frac{1}{\sqrt{P_{succ}}}  \left[\frac{1  -e^{i\frac{2\pi}{3} J_z }}{2}\right]^M\frac{e^{-i\theta J_y} + e^{i\theta J_y}}{2} \ket{+}^{\otimes 9}.
\end{align}
In the noiseless case, setting the number of times you apply the measurement to $M = 5$, gives a fidelity  about $99.9\%$. The probability for obtaining the targeted state $\ket{\varphi_{3,3,1}}$ is $P_{succ}\approx 19.2\%$. 

The preparation described here can be easily generalized for preparing $\ket{\varphi_{g,n,u}}$ with
arbitrary $g,n,u$. The corresponding initial state $\ket{\overline{\varphi}_{g,n,u}}$ now satisfies ($j=0,1,\ldots, n$)
\begin{equation}
\frac{ \braket{N = gnu, m_z = gj -\frac{N}{2}|\overline{\varphi}_{g,n,u} } }{ \braket{N = gnu, m_z = -\frac{N}{2}|\overline{\varphi}_{g,n,u} }   } = \sqrt{\binom{n}{j}}.
\end{equation}
Note that  $\ket{\overline{\varphi}_{g,n,u}}$ can be constructed in the same way as the 9-qubit case in Eq.~(\ref{eq:PI-9QUBIT}), but now we need multiple projectors in the form $\prod_l (e^{-i\theta_l J_y} + e^{i\theta_l J_y})/2$ with different angles $\theta_l$. These angles can also be obtained numerically. 

Similarly, we find that $\ket{\varphi_{g,n,u}}$ is a simultaneous $+1$ eigenstate of the operators
\begin{equation}
    S_{g,n,u}(a) = e^{i\frac{2a\pi}{g}(J_z +\frac{gnu}{2})}
\end{equation} with $a$ integer.
Measuring these operators multiple times with post-selection, qubits in the initial state $\ket{\overline{\varphi}_{g,n,u}}$ can be effectively projected into the target state $\ket{\varphi_{g,n,u}}$.

\section{Adiabatic controlled rotation}\label{sec:adiabatic_JZ}

Here we explain how to apply 
controlled global rotations to a NV ensemble by adiabatically tuning the flux qubit frequency and using the third level of the electronic spin at each NV center. Starting from an appropriate product state, such controlled-rotation can also be used to prepare a highly entangled Dicke state via phase estimation.

For a collection of $N$ identical NV electronic spins which are coupled to a flux qubit, the Hamiltonian can be written in the form
\begin{equation}
\begin{split}
    &H_{\rm sys} = H_0 + H_{\rm coupl},\\
    &H_0 = -\frac{\omega(t)}{2}Z_f  + \Delta\sum_{i=1}^{N} S_{z_i}^2 + W^{\rm ext}\sum_{i=1}^{N}S_{z_i}, \\
    &H_{\rm coupl} =
    -\gamma_e B X_f \otimes \sum_i^{N}\vec{S}_i \cdot \hat{r},
\end{split}
\end{equation}
where $\vec{S}_i=(S_{x_i}, S_{y_i}, S_{z_i})$ is the spin $S=1$ operator for the NV-electronic spin labeled $i$, $Z_f$ and $X_f$ are the Pauli operators of the flux qubit. As we will adiabatically tune the flux qubit frequency $\omega(t)$, it is a function of time $t$. Here the flux qubit is operating at $\epsilon=0$ in Eq.~\eqref{eq:fluxHamiltonian} and the $X_f$-basis is given by two different persistent current states, inducing opposite magnetic fields \cite{orlando, bylander2011noise,clarke2008superconducting}. 

For simplicity, we relabel the three qubit states of the NV electronic spin as
\begin{equation}
\begin{split}
    &\ket{S=1, m_z = +1} = \ket{2}, \\
    &\ket{S=1, m_z = 0} = \ket{0},\\
    &\ket{S=1, m_z = -1} = \ket{1}.
\end{split}
\end{equation}
so that state $\ket{2}$ is outside the computational subspace. Here we assume that $\hat{r}$, the direction orthogonal to the flux loop, is along the $\hat{x}$-direction of the NV-centers. This means that the NV-axis lies in the plane of the flux qubit loop, so that 
the coupling term equals $H_{\rm coupl} =-\gamma_e B X_f \otimes \sum_i^{N} S_{x_i}$. Using the definition $\hat{S}_x = (\ket{2}\bra{0} + \ket{0}\bra{2} + \ket{1}\bra{0} + \ket{0}\bra{1})/\sqrt{2}$, the coupling Hamiltonian can be written as (here we only keep the flip-flop terms)
\begin{equation}
\begin{split}
    H_{\rm coupl} 
    \approx & g \ket{0}\bra{1}_f \otimes \sum_{i=1}^{N}\left( \ket{2}\bra{0}_i + \ket{1}\bra{0}_i \right) +  h.c.,
\end{split}
\end{equation}
where $g = -\gamma_e B/\sqrt{2}$.
The free Hamiltonian $H_0$ can be written as
\begin{equation}
\begin{split}
    H_0 =& -\frac{\omega(t)}{2} \ket{0}\bra{0}_f+\frac{\omega(t)}{2} \ket{1}\bra{1}_f\\ &+ \omega_1\sum_{i=1}^{N} \ket{1}\bra{1}_i + \omega_2\sum_{i=1}^{N} \ket{2}\bra{2}_i,\\
    \omega_1 &= \Delta - W^{ext},\quad \omega_2 = \Delta + W^{ext}.
\end{split}
\end{equation}
With an external magnetic field of $O(100)$ Gauss along the NV axis ($W^{\rm ext} >0$), the frequency difference $\omega_2-\omega_1$ can be as large as $O(1) \mathrm{GHz}$ \cite{abobeih2019imaging}. 
To implement the controlled rotation, we will start from $\omega_1 < \omega(t) < \omega_2$, adiabatically tuning $\omega(t)$ up to $\omega_2$ and then tuning it back. Such adiabatic control can be done by applying a flux through the loop which sets the tunnel barrier of the flux qubit, see, e.g., Ref.~\cite{harris:flux}. If one moves away from the sweet-spot point of this controlling loop, some additional flux noise can be incurred.

We can set $\omega(t)$ to stay far away from $\omega_1$ during the adiabatic path so that we are not activating any flip-flop interactions inside the computational space. 
Hence, we further neglect the off-resonant flip-flop terms between $\ket{1}_{f}\otimes \ket{0}_i$ and $\ket{0}_{f}\otimes \ket{1}_i$ and obtain an approximate interaction Hamiltonian
\begin{equation}
\begin{split}
    \tilde{H}_{\rm coupl} 
    =  & g \sum_{i=1}^N \left[ \ket{0}\bra{1}_f \otimes \ket{2}\bra{0}_i +  \ket{1}\bra{0}_f \otimes \ket{0}\bra{2}_i\right].
\end{split}
\end{equation}

For each Dicke state $\ket{N,m_z}$ (except when $m_z=-N/2$ and all spins are in state $\ket{1}$), we can define a two-dimensional subspace spanned by orthogonal states to which the dynamics is confined:
\begin{equation}
    \begin{split}
        \ket{\phi_0(m_z)} &= \ket{1}_f\otimes\ket{N,m_z},\\
        \ket{\phi_1(m_z)} &= \frac{\tilde{H}_{\rm coupl}\ket{\phi_0(m_z)}}{\sqrt{\bra{\phi_0(m_z)}\tilde{H}_{\rm coupl}^{\dagger}\tilde{H}_{\rm coupl}\ket{\phi_0(m_z)}} }.
    \end{split}
\end{equation}
Note that the states $\ket{\phi_1(m_z)}$ differ from $\ket{\phi_0(m_z)}$ in that one of the NV center qubits in $\ket{0}$ has been flipped to $\ket{2}$, while the flux qubit has been flipped from $\ket{1}$ to $\ket{0}$.
We assume that we start from a product state where the probability of $\abs{m_z} < O(\sqrt{N})$ approaches unity for large $N$, we can neglect the zero effect of the interaction on the state $\ket{N,m_z = -\frac{N}{2}}$. 

In the subspace spanned by $\{\ket{\phi_0(m_z)}, \ket{\phi_1(m_z)} \}$, we write the system Hamiltonian as
\begin{equation}\label{eq:subspaceH}
\begin{split}
    \tilde{H}_{\rm sys} &= H_0 + \tilde{H}_{\rm coupl}\\ 
    &= 	\begin{pmatrix}
        (\frac{N}{2}-m_z)\omega_1+\frac{\omega(t)}{2} & G(m_z) \\
        G(m_z) & (\frac{N}{2}-m_z)\omega_1 + \omega_2 - \frac{\omega(t)}{2}
    \end{pmatrix}
\end{split}
\end{equation}
with effective coupling strength
\begin{equation}
\begin{split}
	G(m_z) &= \sqrt{\bra{\phi_0(m_z)}\tilde{H}_{\rm coupl}^{\dagger}\tilde{H}_{\rm coupl}\ket{\phi_0(m_z)}}\\
	& = g\sqrt{\frac{N}{2}(\frac{N}{2}+1) - m_z(m_z-1)}.
\end{split}
\end{equation}
Note that for the product state $\ket{0\ldots 0}$ with $m_z=N/2$, $G(N/2)=g \sqrt{N}$.
The eigenvalues of $\tilde{H}_{\rm sys}$ in this subspace are 
\begin{equation}
    E(t) = \frac{\omega_2}{2} \pm \sqrt{G(m_z)^2 + \left(\frac{\omega_2-\omega(t)}{2}\right)^2} + \left(\frac{N}{2}-m_z \right)\omega_1.
\end{equation}
On the adiabatic trajectory lasting for time $T$, the following phases can be neglected (or trivially compensated): (\romannumeral1) NV centers qubits in the state $\ket{1}$ which together pick up a total phase $\exp(-i \int_{t=0}^T dt \left(\frac{N}{2}-m_z \right)\omega_1)$; (\romannumeral2) the phase $\frac{1}{2}\omega_2 T$ obtained by the flux qubit being in $\ket{1}_f$; (\romannumeral3) the phase $\exp(i \int_{t=0}^T \frac{\omega(t)}{2}dt) $ obtained by the flux qubit being in $\ket{0}_f$.

The coupling strength $G(m_z)$ is minimum when $m_z=N/2$, and scales as $O(N)$ for $m_z =O(1)$.
Thus, initially when we don't want any interaction, we need to choose $|\omega_2-\omega|\gg G(m_z)$ for all $|m_z| \leq O(\sqrt{N})$, that is, $\omega_2$ and $\omega$ should be sufficiently detuned. With this weak coupling the states $\ket{\phi_0(m_z)}, \ket{\phi_1(m_z)}$ are approximate eigenstates ( besides states such as $\ket{0}_f \ket{N,m_z}$ which do not couple).

The gap on the adiabatic trajectory in the $m_z$-labeled subspace is $\Delta(m_z)=2\sqrt{G(m_z)^2 + \left(\frac{\omega_2-\omega(t)}{2}\right)^2}$ which is minimized at the avoided crossing $\omega(t)=\omega_2$. As we seek to apply this interaction on a state for which $|m_z| \sim O(\sqrt{N})$ and $N$ is large, $\Delta(m_z) \approx gN/2$ at the avoided crossing. We thus assume the path is fully adiabatic.
One could possibly choose a trajectory such as in Ref.~\cite{Martinis2014-zw}.

If we assume that a negligible phase is picked up during a relatively-fast trajectory, followed by a waiting time $\delta t$ at the avoided crossing and a fast switch back, we can see that the state $\ket{\phi_0(m_z)} \rightarrow e^{i \varphi(\delta t)}\ket{\phi_0(m_z)}$ where 
\begin{equation}
\begin{split}
    \varphi(\delta t) & = -G(m_z)\delta t \\
    & \approx g\delta t \left(\frac{m_z^2-m_z}{\sqrt{N(N+2)} } - \frac{\sqrt{N(N+2)}}{2}\right) .
\end{split}
\end{equation}
Here we have assumed that $\abs{m_z} \ll N$ to Taylor expand $G(m_z)$ and neglected higher order terms.
The $m_z$-independent phase $-g\delta t \frac{\sqrt{N(N+2)}}{2}$ can be further compensated by a single-qubit rotation of the flux qubit after the adiabatic trajectory.

Thus in the limit of large spin number $N$, the adiabatic procedure approximately applies the unitary
\begin{equation}\label{eq:adiabatic_controlled_rotation}
V(\delta t) = \ket{0}\bra{0}_f \otimes I + \ket{1}\bra{1}_f\otimes \exp{\left(\frac{i g\delta t}{N} (J_z^2 -J_z)  \right)  }.
\end{equation}
Now we will start the preparation from a product state $e^{-i2\chi J_y}\ket{+}^{\otimes N}$ with an appropriate value of $\chi$ as in Eq.~(\ref{eq:rotated_intial_superposition}). Here we consider a specific example: we choose the value of $\chi$ so that the product state $e^{-i2\chi J_y}\ket{+}^{\otimes N}$ is a superposition of Dicke states, which are localized around $\ket{N, m_z= 2\sqrt{N}}$. When $N$ is large, the probability for $0 \leq m_z \leq 4\sqrt{N}$ approaches unity. For this product state, we can make an approximation that the eigenvalues of $J_z$ and $J_z^2-J_z$ are in 1-1 correspondence. We can use this form of controlled rotation in Eq.~(\ref{eq:adiabatic_controlled_rotation}) to perform phase estimation and then prepare $\ket{N,m_z\sim O(\sqrt{N})}$. 

However, this adiabatic form of applying a controlled rotation makes it hard to get a strong interaction, as the rotation angle scales as $O(1/N)$.

\bibliography{main.bib}

\end{document}